\pdfoutput=1

\documentclass{article}

\usepackage[final]{neurips_2021}

\usepackage[hang,flushmargin]{footmisc}

\usepackage[utf8]{inputenc} 
\usepackage[T1]{fontenc}    
\usepackage[hidelinks]{hyperref}       
\usepackage{url}            
\usepackage{booktabs}       
\usepackage{amsfonts}       
\usepackage{amssymb}
\usepackage{nicefrac}       
\usepackage{microtype}      
\usepackage{xcolor}         
\usepackage{amsmath}
\usepackage{xspace}
\usepackage{enumitem}
\usepackage{arydshln}

\newcommand{\cls}[0]{\texttt{[CLS]}\xspace}

\title{A Proposed Conceptual Framework for a Representational Approach to Information Retrieval}

\author{
  Jimmy Lin \\ [1ex]
  David R. Cheriton School of Computer Science \\
  University of Waterloo
}

\begin{document}

\maketitle

\begin{abstract}
This paper outlines a conceptual framework for understanding recent developments in information retrieval and natural language processing that attempts to integrate dense and sparse retrieval methods.
I propose a representational approach that breaks the core text retrieval problem into a logical scoring model and a physical retrieval model.
The scoring model is defined in terms of encoders, which map queries and documents into a representational space, and a comparison function that computes query--document scores.
The physical retrieval model defines how a system produces the top-$k$ scoring documents from an arbitrarily large corpus with respect to a query.
The scoring model can be further analyzed along two dimensions:\ dense vs.\ sparse representations and supervised (learned) vs.\ unsupervised approaches.
I show that many recently proposed retrieval methods, including multi-stage ranking designs, can be seen as different parameterizations in this framework, and that a unified view suggests a number of open research questions, providing a roadmap for future work.
As a bonus, this conceptual framework establishes connections to sentence similarity tasks in natural language processing and information access ``technologies'' prior to the dawn of computing.

\end{abstract}

\section{Introduction}

For the past half a century, information retrieval has been dominated by bag-of-words exact-match scoring models such as BM25 executed at scale using inverted indexes and efficient query-at-a-time retrieval algorithms.
Even in the context of feature-based learning to rank and, more recently, neural models, these bag-of-words models remain of fundamental importance because they provide potentially relevant texts for downstream reranking in the context of multi-stage pipelines.
This role is usually referred to as first-stage retrieval or candidate generation.
Multi-stage ranking architectures have been studied extensively by academic researchers~\citep{Matveeva_etal_SIGIR2006,Cambazoglu_etal_WSDM2010,Wang_etal_SIGIR2011,Tonellotto_etal_WSDM2013,Asadi_Lin_SIGIR2013,Capannini_etal_IPM2016,Clarke_etal_IRJ2016,ChenRuey-Cheng_etal_SIGIR2017a,Mackenzie_etal_WSDM2018} and there is substantial documentation that many commercial applications are designed in this manner~\citep{Pedersen_SIGIR2010,LiuShichen_etal_SIGKDD2017,HuangJui-Ting_etal_SIGKDD2020,Zou:2105.11108:2021}.

There has, of late, been much interest and excitement surrounding so-called ``dense retrieval'' techniques, or ranking with learned dense representations.
This general approach, often called a bi-encoder design~\citep{Humeau_etal_ICLR2020}, is perhaps best exemplified by DPR~\citep{karpukhin-etal-2020-dense} and ANCE~\citep{Xiong_etal_ICLR2021}, but other examples abound~\citep{Gao_etal_ECIR2021_CLEAR,Hofstatter:2010.02666:2020,qu-etal-2021-rocketqa,Hofstatter_etal_SIGIR2021,qu-etal-2021-rocketqa,Zhan_etal_SIGIR2021,Lin_etal_2021_RepL4NLP}.
Dense retrieval is formulated as a representational learning problem where the task is to learn (nowadays, transformer-based) encoders that map queries and documents into dense fixed-width vectors (768 dimensions is typical).
The goal is to maximize inner products between queries and relevant documents and to minimize inner products between queries and non-relevant documents.
This is framed as a supervised machine learning problem, with relevance signals coming from a large dataset such as the MS MARCO passage ranking test collection~\citep{MS_MARCO_v3}.
\citet{Lin_etal_2021_ptr4tr} provide a recent survey of this general approach within the broader context of text ranking using BERT and other pretrained transformer-based language models.

Experiments have shown that dense retrieval methods outperform ``sparse retrieval'' methods, usually referring to bag-of-words exact-match methods such as BM25.\footnote{Referring to bag-of-words exact-match methods as ``sparse retrieval'' is a relatively new invention, primarily to establish contrast with dense retrieval methods. Nevertheless, I will use this terminology throughout the paper.}
This appears to be a robust and widely replicated finding, and dense retrieval models are known to have been deployed in real-world search applications, for example, by Bing~\citep{Xiong_etal_ICLR2021} and Facebook~\citep{HuangJui-Ting_etal_SIGKDD2020}.
Scaling such methods requires infrastructure that is very different from sparse retrieval:\ instead of relying on inverted indexes for query evaluation, as BM25 does, dense retrieval typically relies on approximate nearest neighbor (ANN) search; one standard technique exploits hierarchical navigable small world graphs (HNSW)~\citep{Malkov_Yashunin_2020}.

Thus, recent literature appears to have established a contrast between dense retrieval and sparse retrieval.
The standard portrayal is that they represent fundamentally different approaches, requiring different problem formulations, different models, and different software infrastructures for efficient execution at scale.
I argue, however, that this is not the case.
Aspects of the ideas and observations presented here were originally captured in two previous papers~\citep{Lin_Ma_arXiv2021,Lin_etal_DESIRES2021}.
I build on both, with additional analysis and synthesis.

The goal of this paper is to provide a conceptual framework that unites dense and sparse retrieval by demonstrating that they, in fact, have the same functional form, just with different parameterizations.
This framework adopts a representational approach and breaks the core text retrieval problem into a logical scoring model and a physical retrieval model, allowing a researcher to separate how document relevance scores are computed from how retrieval is performed at scale.
In terms of scoring models, dense and sparse retrieval can be characterized along two dimensions:\ the contrast between dense vs.\ sparse vector representations, and the contrast between supervised (learned) vs.\ unsupervised approaches.

The main contribution of this conceptual framework is that it provides abstractions to help researchers make sense of the panoply of recently proposed retrieval models that, at first glance, defy orderly categorization.
The proposed framework suggests a number of open research questions, providing a roadmap for future research, potentially tying together multiple sub-fields within information retrieval.
As a bonus, this conceptual framework establishes interesting connections to sentence similarity tasks in natural language processing and information access ``technologies'' prior to the dawn of computing.

\section{A Conceptual Framework}

The formulation of text retrieval (alternatively, text ranking)---what information retrieval researchers more precisely call {\it ad hoc} retrieval---is typically defined as follows:\
Given an information need expressed as a query $q$, the text retrieval task is to return a ranked list of $k$ documents\footnote{Consistent with parlance in information retrieval, I use ``document'' throughout this paper in a generic sense to refer to the unit of retrieved text, even though in truth it may be a passage, a web page, a PDF, or some arbitrary span of text.} $\{d_1, d_2 \ldots d_k\}$ from an arbitrarily large but finite collection of documents $\mathcal{D} = \{ d_i \}$ that maximizes a metric of interest, for example, nDCG, AP, etc.
These metrics vary, but they all aim to quantify the ``goodness'' of the results with respect to the information need; in some cases, metrics can be understood more formally in terms of the utility that a user would derive from consuming the results.
The retrieval task is also called top-$k$ retrieval (or ranking), where $k$ is the length of the ranked list (also known as the retrieval or ranking depth).

We can break the text retrieval problem down into two distinct components, as follows:

\paragraph{Logical Scoring Model}

Let us define $\eta_q(q)$ and $\eta_d(d)$ as two arbitrary functions that take a query and a document (both sequences of terms), respectively, and map each into a fixed-width vector representation.
As will become clear below, I will call these two functions ``encoders''.

Let us further define a comparison function $\phi$ that takes these fixed-width vector representations and computes a score.
We have:
\begin{equation}
s(q, d) \overset{\Delta}{=} \phi(\eta_q(q), \eta_d(d))
\label{eq:logical}
\end{equation}

\noindent We can interpret the score $s$ as quantifying the degree to which $d$ is relevant to query $q$, i.e., the basis for ranking a set of documents with respect to a query.
For example, we desire to maximize scores for queries and their relevant documents and minimize scores for queries and non-relevant documents (note how this statement can be straightforwardly operationalized into a loss function).
For dense retrieval methods, this design is commonly called a bi-encoder~\citep{Humeau_etal_ICLR2020}.

More intuitively, we can understand the score $s$ as capturing the probability of relevance:
\begin{equation}
P(\textrm{Relevant}=1 | d, q) \overset{\Delta}{=} s(q, d).
\label{eq:prob-rel}
\end{equation}

\noindent Note that the domain of $\eta_q$ comprises arbitrary sequences of terms, including sequences that have never been encountered before.
In contrast, the domain of $\eta_d$ is typically $\mathcal{D}$, since we are retrieving from a given collection of documents (i.e., the corpus).

The logical scoring model, as defined in Eq.~(\ref{eq:logical}), nicely captures why I characterize this proposed conceptual framework as a ``representational approach'', since it focuses on matching {\it representations} derived from queries (information needs) and documents (texts to be searched).
In the context of bag-of-words representations, this formulation puts the vocabulary mismatch problem~\citep{Furnas87}---overcoming the fact that information seekers and authors use different words to express the same concepts---front and center in the design of retrieval models.
As I will discuss in detail later, neural models are simply the source of (better) representations---the structure of the {\it ad hoc} retrieval problem remains the same.
In fact, across many diverse formulations of retrieval models, $\phi$ is defined as the inner product.

\paragraph{Physical Retrieval Model}
Given the setup above, top-$k$ retrieval can be defined as:
\begin{equation}
\underset{d \in \mathcal{D}}{\textrm{arg top-k }} \phi(\eta_q(q), \eta_d(d))
\label{eq:physical}
\end{equation}
\noindent That is, given $q$, we wish to identify from $\mathcal{D}$ the $k$ documents $d_1 \ldots d_k$ that have the highest scores $s_1 \ldots s_k$.
These $\{ (d_i, s_i) \}_{i=0}^{k}$ pairs are usually referred to as the ranked list of results (sometimes called the ``hits'').

If $s$ is interpreted as a probability of relevance, as per Eq.~(\ref{eq:prob-rel}), then the physical retrieval model represents a direct realization of the \textit{Probability Ranking Principle}~\citep{Robertson77}, which states that documents should be ranked in decreasing order of the estimated probability of relevance with respect to the query.

\bigskip
\noindent We might think of the logical scoring model and the physical retrieval model as providing what I argue to be the ``right'' abstractions for the text retrieval problem.
So far, however, nothing in the presentation above captures information that isn't already common knowledge.
I have simply adopted notation that may seem slightly peculiar, compared to how the text retrieval problem is usually presented (for example, in standard textbooks).
Nevertheless, I will attempt to convince the reader that this isn't a pointless symbol manipulation exercise, but rather this framing of the problem provides a conceptual framework that bridges dense and sparse retrieval methods.

\subsection{Applications to Dense and Sparse Retrieval}
\label{section:applications}

Let us consider DPR~\citep{karpukhin-etal-2020-dense}, a popular representative dense retrieval model, and see how it can be understood within this conceptual framework.
DPR uses separate transformer-based encoders for queries and documents, $\eta_q$ and $\eta_d$, respectively.
Both encoders take the \cls representation from BERT~\citep{devlin-etal-2019-bert} as its output representation.
In other words, the DPR encoders project queries and documents into fixed-width vector representations in some latent semantic space (by default, 768 dimensions).

Relevance between query representations and document representations---the comparison function $\phi$---is defined in terms of inner products:
\begin{equation}
\phi(\eta_q(q), \eta_d(d)) = \eta_q(q)^\intercal \eta_d(d)
\end{equation}

\noindent The model is trained as follows:\ let $\mathcal{R} = \{ \langle q_i, d_i^+, d_{i,1}^-, d_{i,2}^-, \ldots d_{i,n}^- \rangle \}_{i=1}^m$ be the training set comprising $m$ instances.
Each instance contains a query $q$, a relevant passage $d^+$, and $n$ non-relevant passages $d_1^-, d_2^-,...d_n^-$.
DPR is trained with the following loss function:
\begin{equation}\label{eq:dpr_obj}
\mathcal{L}(q, d^+, d_1^-, d_2^-,...d_n^-) = - \log \frac{ \exp \left[ \phi( \eta_q(q),  \eta_d(d^+))\right] }{\exp \left[ \phi(\eta_q(q), \eta_d(d^+)) \right] + \sum_{j=1}^{n} \exp \left[ \phi(\eta_q(q), \eta_d(d_j^-)) \right]}.
\end{equation}

\noindent Non-relevant passages for a query are selected via in-batch negative sampling~\citep{Henderson:1705.00652:2017}, from examples associated with other queries in the same training batch.
However, this is a technical detail and other models select negative examples in different ways.
For example, ANCE~\citep{Xiong_etal_ICLR2021} searches for ``hard negatives'' based on an earlier version of the document encoder itself.

I have just described DPR in terms of the proposed conceptual framework outlined above.
Now let's try to recast BM25~\citep{Robertson95} in the same framework.
In fact, the mapping is pretty straightforward:
The query encoder $\eta_q$ and the document encoder $\eta_d$ both generate sparse bag-of-words vector representations of dimension $|V|$, where $V$ is the vocabulary of the corpus.
For the output of the document encoder $\eta_d$, as with any bag-of-words representation, each dimension corresponds to a term in the vocabulary, and each term is assigned a weight according to the BM25 scoring function.
The query encoder $\eta_q$ uses a multi-hot representation, with a weight of one if the term is present in the query, and zero otherwise.\footnote{This is a slight simplification; the original formulation of BM25~\citep{Robertson95} included a query weighting component, but this term is usually omitted in modern implementations~\citep{Kamphuis_etal_ECIR2020}.}
The comparison function $\phi$ is, like DPR, defined in terms of the inner product.

Viewed in this manner, we can clearly see that BM25 and DPR have the same functional form, parameterized by $\eta_q$, $\eta_d$, and $\phi$, and in fact, $\phi$ is the inner product in both cases.
Explained in terms of abstractions such as interfaces in programming languages, by analogy the logical scoring model defines the abstract methods ($\eta_q$, $\eta_d$, and $\phi$) that specific retrieval models override with custom implementations, and here I have demonstrated that the abstraction covers both BM25 and DPR.
This framework can be applied to the recent panoply of proposed dense retrieval methods in the literature, as well as nearly all families of bag-of-words exact-match models beyond BM25's probabilistic formulation, e.g., tf--idf, query likelihood, divergence from randomness, etc.
This conceptual framework allows us to draw a direct connection between dense retrieval and sparse retrieval as parametric variations of the same underlying logical scoring model.

Finally, what about cross-encoders?
Typical of this design is the monoBERT model~\citep{nogueira2019passage,Lin_etal_2021_ptr4tr}, where a query and a document are fed into a pretrained transformer as part of an input template, and the contextual representation of the {\small \texttt{[CLS]}} token is used for relevance classification.
Here, we can say that the comparison function $\phi$ is defined in terms of the transformer, and thus cross-encoders are still captured by the logical scoring model defined in Eq.~(\ref{eq:logical}).

``Hiding'' transformer inference in the comparison function $\phi$ might seem like a sleight of hand, but the PreTTR reranking model proposed by \citet{MacAvaney_etal_SIGIR2020a} connects a ``full'' cross-encoder like monoBERT on the one hand to $\phi$-as-inner-product methods like DPR on the other hand.
MacAvaney et al.~began with the simple observation that query--document attention prevents document representations from being computed offline; recall that in DPR, $\eta_d(\cdot)$ does not depend on the query.
Yet, it is precisely query--document attention that allows cross-encoders to obtain high levels of effectiveness.
PreTTR was designed with this insight:\ What if we limited query--document attention to only the upper layers of the transformer?
In such a design, document representations in the lower layers could be precomputed (and hence cached to accelerate inference).
At one extreme end of the PreTTR design space, if {\it all} query--document attention is eliminated, then we have essentially ``cleaved'' monoBERT into two disconnected networks, and the result looks quite similar to DPR, where each of the disconnected networks serves as an encoder (and all document representations can be precomputed and indexed for low-latency retrieval).
At the other extreme, if no query--document attention is eliminated, we have monoBERT.
Thus, PreTTR provides the conceptual linkage that allows us to understand bi-encoders and cross-encoders as the two extreme cases of a single underlying design:\ it's all in the definition of the comparison function $\phi$.

\subsection{Generalization of Logical Scoring Models}
\label{section:logical}

Dense retrieval models such as DPR are often compared against sparse retrieval models such as BM25 in experimental evaluations, as~\citet{karpukhin-etal-2020-dense} did in their paper.
Not surprisingly, results show that dense retrieval models obtain higher effectiveness.

This, however, is not a fair comparison.
Dense retrieval methods represent an instance of representational learning---the key here is {\it learning}.
The output of the encoders are {\it learned} representations that benefit from (large amounts of) training data under a standard supervised machine learning paradigm.
In contrast, BM25 is {\it unsupervised}.\footnote{Leaving aside simple tuning of parameters such as $k_1$ and $b$.}
Comparing a supervised method to an unsupervised method is fundamentally an apples-to-oranges juxtaposition; it should not be surprising that a supervised technique is more effective.

\begin{table}[t]
\centering
\begin{small}
\begin{tabular}{l|l|l}
&  {\bf Dense} & {\bf Sparse} \\
\hline
{\bf Supervised} & DPR, ANCE & DeepImpact, uniCOIL \\
{\bf Unsupervised} & LSI, LDA & BM25, tf--idf \\
\end{tabular}
\end{small}
\vspace{0.2cm}
\caption{A taxonomy of logical scoring models.}
\label{table:framework}
\end{table}

As previously argued in~\citet{Lin_Ma_arXiv2021}, the encoders $\eta_{\cdot}$ should be organized along two distinct dimensions or properties:
The first dimension contrasts dense vs.\ sparse vector representations for queries and documents.
The second dimension distinguishes between supervised (learned) and unsupervised representations.
Table~\ref{table:framework} illustrates this taxonomy.
DPR (along with nearly all dense retrieval methods today) are instances of learned dense representations.
BM25 is an instance of an unsupervised sparse representation.

This taxonomy immediately points to the existence of two other classes of logical scoring models.
In fact, they correspond to models described in the literature that we can now categorize and unify in a single conceptual framework:

\paragraph{Learned sparse representations}
The existence of learned dense representations such as DPR and unsupervised sparse representations such as BM25 suggests that there should exist a class of learned sparse representations.

Learning sparse representations is by no means a new idea.
If we fix the dimensions of the output representation to be the vocabulary (i.e., retaining a bag-of-words assumption), models for learned sparse representations become term weighting models---that is, a supervised machine learning approach to learning term weights.
The earliest example I am aware of is~\citet{Gordon_CACM1988}, who applied (what we might today call) representational learning on boolean vectors of descriptors using genetic algorithms, based on a small set of relevance judgments.
These experiments might today be characterized as ``toy'', but all the key elements of learned sparse retrieval models (quite amazingly!)~are present.
Another example along these lines is the work of~\citet{Wilbur_2001}, who attempted to learn global term weights using TREC data.
A bit later, \citet{Trotman_IRJ2005} used genetic programming to discover better BM25-like scoring functions.
Quite simply, there is plenty of evidence that learned sparse representations aren't new.

The first example of learned sparse representations in the ``BERT era'' is DeepCT~\citep{Dai:1910.10687:2019}, which uses a transformer to learn term weights based on a regression model, with the supervision signal coming from the MS MARCO passage ranking test collection.
DeepCT has an interesting ``quirk'':\ in truth, it only learns the term frequency (tf) component of term weights, but still relies on the remaining parts of the BM25 scoring function via the generation of pseudo-documents.
The method also has a weakness:\ it only assigns weights to terms that are already present in the document, which limits retrieval to exact match.
More generally, if we retain a bag-of-words assumption, term weighting models cannot address the vocabulary mismatch problem (more below).
Note that dense representations do not have this issue since the dimensions of the vector representation capture some latent semantic space, not specific terms in the corpus vocabulary, and thus are able to capture what researchers call ``semantic matching''.

The exact-match weakness of DeepCT discussed above was resolved by the DeepImpact model \citep{Mallia_etal_SIGIR2021}, which brought together two key ideas:\
the use of document expansion to identify dimensions in the sparse bag-of-words representation that should have non-zero weights and a term weighting model based on a pairwise loss between relevant and non-relevant documents with respect to a query.
Expansion terms are identified by doc2query--T5~\citep{Nogueira_Lin_docTTTTTquery}, a sequence-to-sequence model for document expansion that predicts queries for which a text would be relevant.
Since DeepImpact directly predicts term weights that are then quantized, it would be more accurate to call these weights learned impacts, since query--document scores are simply the sum of weights of document terms that are found in the query.
Furthermore, calling these impact scores draws an explicit connection to a thread of research in information retrieval dating back two decades~\citep{Anh_etal_SIGIR2001}.

Many other retrieval models can also be understood as instances of learned sparse representations, which allow for different parameterizations.
\citet{Lin_Ma_arXiv2021} argued that another recent model called COIL~\citep{gao-etal-2021-coil} is an instance of learned sparse representations, where the scoring model assigns each term a {\it vector} ``weight'', stored in standard inverted lists.
Lin and Ma demonstrated this connection by introducing a degenerate version of COIL called uniCOIL, where the weight vectors are collapsed down into a single dimension, thus yielding {\it scalar} weights.

In this proposed conceptual framework, we might implement document expansion differently:\ uniCOIL originally used doc2query--T5 for document expansion, but this was replaced by \citet{Zhuang:2108.08513:2021} with an alternative model based on TILDE~\citep{Zhuang_Zuccon_SIGIR2021}.
They demonstrated that expansion using TILDE achieves comparable effectiveness on the MS MARCO passage ranking task, but with substantially lower inference costs.
As another interesting variation, note that the query and document encoders need not be based on transformers (e.g., \citet{Zamani_etal_CIKM2018}), or even neural networks at all!
For example, the retrieval model of \citet{Boytsov:2010.14848:2020}, which exploits translation probabilities learned from query--passage pairs, can be considered a (non-neural) learned sparse model.

Synthesizing recent literature, there are three important observations about retrieval using learned sparse representations, which were originally noted by~\citet{Lin_Ma_arXiv2021}:

\begin{itemize}[leftmargin=0.75cm]

\item {\it Choice of basis.}
When contrasting learned dense representations with learned sparse representations, we see that nearly all recent proposals take advantage of transformers (\citet {Boytsov:2010.14848:2020} being a notable exception), so that aspect of the design is not a salient distinction.
The critical difference is the basis of the vector representations:\
In nearly all current sparse approaches, the basis of the vector space remains fixed to the corpus vocabulary, i.e., they retain the bag-of-words assumption, even though in principle one could imagine sparse representations that abandon this assumption.
In dense approaches, the model is given the freedom to ``choose'' a new basis derived from transformer representations.
This change in basis allows the encoder to represent the ``meaning'' of texts in relatively small fixed-width vectors (say, 768 dimensions, compared to sparse vectors that may have millions of dimensions).
This leads us to the next important observation:

\item {\it Expansions for sparse representations.}
Without some form of expansion, learned sparse representations remain limited to (better) exact matching between queries and documents.
The nature of sparse representations means that it is computationally impractical to consider non-zero weights for {\it all} elements in the vector (i.e., the vocabulary space).
Thus, document expansion serves the critical role of proposing a set of candidate terms that {\it should} receive non-zero weights; since the number of candidate terms is small compared to the vocabulary size, the resulting vector remains sparse.
Without some form of expansion, learned sparse representations cannot address the vocabulary mismatch problem~\citep{Furnas87}, because document terms not present in the query cannot contribute any score.
This leads us to the third important observation:

\item {\it Expansion and Term Weighting.}
The upshot of the above analysis is that retrieval methods based on learned sparse representations can be decomposed into an expansion and a term weighting component.
For example, DeepCT performs no expansion and uses a regression-based scoring model.
DeepImpact performs document expansion with doc2query--T5, and as discussed above, the doc2query--T5 model can be replaced with the TILDE document expansion model~\citep{Zhuang:2108.08513:2021}.
Although many learned sparse models today have distinct expansion and weighting components, one can certainly imagine an integrated end-to-end model that jointly performs both.
Nevertheless, such models will still need to tackle these distinct challenges:\ overcoming vocabulary mismatch and predicting term importance.

\end{itemize}

\noindent I will examine the impact of different design decisions for learned sparse representations in Section~\ref{section:exp}, drawing on recent experimental results from the literature.

\paragraph{Unsupervised dense representations.}
The juxtaposition of DPR and BM25 suggests the existence of learned sparse representations.
Establishing dense vs.\ sparse and supervised (learned) vs.\ unsupervised as the relevant dimensions of contrast suggests a class of unsupervised dense methods.
While there is little work in this space of late, this label does describe techniques such as LSI~\citep{Deerwester_etal_1990,Atreya_Elkan_2010} and LDA~\citep{Wei06}, which have been previously explored.
I don't have much to say here, except that perhaps this gap might highlight a research direction worth renewed investigation.

\bigskip
\noindent Based on this discussion, we see that all quadrants in the taxonomy of logical scoring models shown in Table~\ref{table:framework} are populated with known examples from the literature.
Furthermore, I demonstrate (hopefully, in a convincing manner) that all of these methods can be viewed as different $\eta_q$, $\eta_d$, and $\phi$ parameterizations of the logical scoring model captured in Eq.~(\ref{eq:logical}).

\subsection{Logical/Physical Separation}
\label{section:logical-physical}

The logical scoring model in Eq.~(\ref{eq:logical}) describes how query--document scores are to be computed with respect to an arbitrary (query, document) pair.
The text retrieval problem, however, requires a system to produce a top-$k$ ranking from an arbitrarily large collection of documents; this is the goal of what I've called the physical retrieval model, Eq.~(\ref{eq:physical}).
In other words, the end-to-end problem requires the execution of the logical scoring model at scale.

The simplest physical retrieval model is to brute-force compute, given a query, the query--document score for {\it every} document in the collection.
In fact, for research experiments, this remains a common approach for dense retrieval methods, for example, using so-called ``flat'' indexes in Facebook's Faiss library~\citep{faiss}.
For sparse retrieval, in the early days of information retrieval prior to the development of inverted indexes and associated query evaluation algorithms (see~\cite{Perry_Willett_1983}), this was also a common approach.

Obviously, a brute-force scan of sizeable collections is impractical for low-latency querying, with the exception of a few specialized cases~\citep{Lempel_etal_CIKM2007,Wang_Lin_ICTIR2015}.
For dense vector representations, the top-$k$ retrieval problem is often called nearest neighbor (NN) search, and for a small set of $\phi$ comparison functions (inner products, L1 distance, and a few others), there exist efficient, scalable solutions.
This problem has been studied for over two decades, with early solutions relying on locality-sensitive hashing~\citep{Indyk_Motwani_1998,Gionis_etal_VLDB1999}.
Recently, approaches based on hierarchical navigable small-world graphs (HNSW)~\citep{Malkov_Yashunin_2020} have emerged as the preferred solution, and are implemented in a variety of open-source libraries.
Note that these techniques solve the {\it approximate} nearest neighbor (NN) search problem, which means that the top-$k$ they generate are not exact; see, for example,~\cite{Indyk_Motwani_1998} for how this approximation is typically formalized.

For sparse retrieval, nearly all models adopt the inner product as the comparison function $\phi$, and the top-$k$ retrieval problem is solved using efficient query evaluation algorithms (mostly document-at-a-time techniques) operating over inverted indexes.
There has, literally, been decades of work on efficient implementations; see~\cite{Tonellotto_etal_FnTIR2018} for a survey.

With respect to the design of physical retrieval models, there are two important points worth explicitly discussing:

\begin{itemize}[leftmargin=0.75cm]

\item {\it Defining $\phi$ as inner products.}
Although the comparison function $\phi$ can be arbitrarily defined in the logical scoring model, for both dense and sparse representations, defining $\phi$ in terms of inner products (and a small number of other functions) leads to efficient scalable solutions for the top-$k$ retrieval problem.
That is, an inner product formulation of $\phi$ is privileged or ``special''.
If a researcher fixes $\phi$ to be the inner product and only redefines $\eta_q$ and $\eta_d$ to create a new logical scoring model, then existing software infrastructure for efficient top-$k$ retrieval (implemented in various software libraries) can be reused.
In the sparse retrieval space, the development of different scoring models such as tf--idf, BM25, query-likelihood, divergence from randomness, etc., can be characterized as such, as well as most recent work in the dense retrieval space.
In other words, efficient physical retrieval comes ``for free''.

\item {\it Tight logical/physical coupling.}
The current state of affairs can be characterized as follows:\ for sparse representations, top-$k$ retrieval is almost always performed using inverted indexes, typically with document-at-a-time scoring.
For dense representations, the same role is usually filled by HNSW, implemented in Faiss or some other toolkit.
In other words, we observe tight coupling between the logical scoring model and the physical retrieval model.
Thus, dense and sparse representations use completely different ``software stacks''.

\end{itemize}

\noindent The separation of the physical retrieval model from the logical scoring model espoused in this paper represents an explicit attempt to move away from the tight coupling discussed above.
Why can't we perform nearest neighbor search using inverted indexes?
Similarly, why can't we perform BM25 retrieval using HNSW?
There is no reason why not, and in fact, both have already been tried!
\citet{Teofili_Lin_arXiv2019} evaluated a number of ``tricks'' for performing top-$k$ ranking on dense vectors with inverted indexes using the open-source Lucene search library.
\citet{Tu_etal_ICTIR2020} and \citet{Lin_etal_DESIRES2021} explored using HNSW for BM25 ranking.

As it turns out, dense retrieval using inverted indexes doesn't work very well, and sparse retrieval using HNSW appears to be attractive only in limited settings.
In terms of both efficiency and effectiveness, using the ``other'' physical technique to execute the logical scoring model is worse than its ``natural'' counterpart.
Thus, it might be fair to say that sparse representations have an affinity with inverted indexes and dense representations with HNSW.
While possible in principle, there doesn't seem to be a compelling case at present to adopt a decoupled approach.
So what's the point?

At a high level, tight coupling presents optimizations opportunities, while loose coupling promotes flexibility---and I argue that this is exactly what's happened here.
Over the course of many decades, researchers have devised numerous optimizations specifically targeted at efficient query evaluation using inverted indexes for sparse retrieval models~\citep{Tonellotto_etal_FnTIR2018}.
Thus, it is entirely believable (and perhaps even expected) that HNSW---a much newer technique that has received far less attention---cannot compete.
However, it is also plausible that as HNSW receives more attention for different use cases and hence more optimization efforts over time, the performance gap closes.
Explicitly promoting logical/physical separation in a loosely-coupled approach, I argue, increases the range of usage scenarios in which HNSW (and future techniques) may be applied, and thus might hasten these developments.

Even more interesting to consider are representations that are not really dense, but not sparse either.
For such a design, the ability to ``mix and match'' logical scoring models and physical retrieval models presents an interesting future direction.
I come back to discuss this point in more detail in Section~\ref{section:discussion}.

The other major benefit of the logical/physical separation is that it allows us to understand multi-stage ranking as {\it practical} physical realizations of expensive logical scoring models.
For example, in Section~\ref{section:applications}, I argued that cross-encoders like monoBERT are covered by the functional form presented in Eq.~(\ref{eq:logical}), where the comparison function $\phi$ is defined in terms of transformers.
Due to query--document attention, the monoBERT logical scoring model can only be faithfully realized by computing the scores of all $(q, d)$ pairs, $\forall d \in \mathcal{D}$.
This is obviously impractical, and thus one solution to the physical retrieval problem is to adopt a multi-stage design with a ``cheap'' first-stage retrieval.\footnote{Using bag-of-words (unsupervised) sparse retrieval, with $\phi$ defined in terms of the inner product, no less!}
It seems a bit silly to phrase as follows, given the obviousness and triviality of the observation, but defining $\phi$ in terms of transformers does not admit an efficient top-$k$ retrieval solution over large corpora.
The transformer is not one of those privileged functional forms of $\phi$ discussed above.

Supporting evidence for this view comes from an experimental result presented in \citet{Lin_etal_2021_ptr4tr} (Section 3.2.2), who began with a standard BM25 + monoBERT reranking design~\citep{nogueira2019passage} and successively increased the reranking depth.
They performed experiments that applied monoBERT to rerank increasingly larger candidate sets from first-stage retrieval on the MS MARCO passage corpus.
On the associated passage ranking task, Lin et al.~discovered that effectiveness increases (and then plateaus) as the reranking depth increases, out to 50k hits per query.
Given the resource requirements of such an experiment, the authors did not increase reranking depth any further.

These results can be interpreted as follows:\ As the reranking depth increases, the final ranking becomes increasingly closer to a brute-force scan over the entire collection (and, critically, in this method, the final ranking score does {\it not} take into account the BM25 retrieval score).
This interpretation is consistent with the arguments I made above.
To be more precise, multi-stage ranking is an {\it approximate} physical retrieval realization of the monoBERT logical scoring model, since empirically, a smaller $k$ in first-stage top-$k$ retrieval degrades effectiveness.
In the limit, if $k = |\mathcal{D}|$, then we're back to a brute-force computation of query--document scores for all documents in the collection.

So, in summary, decoupling the logical scoring model from the physical retrieval model offers two conceptual advances:\ unifying retrieval with dense and sparse representations, and providing a new perspective for understanding multi-stage ranking.

\subsection{Connections to Natural Language Processing}

\citet{Lin_etal_2021_ptr4tr} argued that relevance, semantic equivalence, paraphrase, entailment, and a host of other ``sentence similarity'' tasks are all closely related, even though the first is considered an IR problem and the remainder are considered to be problems in NLP.
What's the connection?
Cast in terms of the conceptual framework proposed in this paper, I argue that these problems all share in the formalization of the logical scoring model, but NLP researchers usually don't care about the physical retrieval model.

For example, supervised paraphrase detection is typically formalized as a ``pointwise'' estimation task of the ``paraphrase relation'':
\begin{equation}
P(\textrm{Paraphrase}=1 | s_1, s_2) \overset{\Delta}{=} r(s_1, s_2).
\end{equation}
That is, the task is to induce some scoring function based on training data that provides an estimate of the likelihood that two texts (sentences in most cases) are paraphrases of each other.
In the popular transformer-based Sentence-BERT model~\citep{reimers-gurevych-2019-sentence}, the solution is formulated in a bi-encoder design:
\begin{equation}
r(s_1, s_2) \overset{\Delta}{=} \phi(\eta(s_1), \eta(s_2)),
\end{equation}
which has exactly the same functional form as the logical scoring model in Eq.~(\ref{eq:logical})!
The main difference, I argue, is that paraphrase detection for the most part does not care where the texts come from.
In other words, there isn't an explicitly defined physical retrieval model.

In fact, comparing Sentence-BERT with DPR, we can see that although the former focuses on sentence similarity tasks and the latter on passage retrieval, the functional forms of the solutions are identical.
Both are captured by the logical scoring model in Eq.~(\ref{eq:logical}); the definitions of the encoders are also quite similar, both based on BERT, but they extract the final representations in slightly different ways.
Of course, since DPR was designed for a question answering task, the complete solution requires defining a physical retrieval model, which is not explicitly present in Sentence-BERT.

Pursuing these connections further, note that there are usage scenarios in which a logical scoring model for paraphrase detection might require a physical retrieval model.
Consider a community question answering application~\citep{Srba_Bielikova_2016}, where the task is to retrieve from a knowledge base of (question, answer) pairs the top-$k$ questions that are the closest paraphrases of a user's question.
Here, there would be few substantive differences between a solution based on Sentence-BERT and DPR, just slightly different definitions of the encoders.

One immediate objection to this treatment is that relevance differs from semantic equivalence, paraphrase, entailment, and other sentence similarity tasks in fundamental ways.
For example, the relations captured by sentence similarity tasks are often symmetric (with entailment being an obvious exception), i.e., $r(s_1, s_2) = r(s_2, s_1)$, while relevance clearly is not.
Furthermore, queries are typically much shorter than their relevant documents (and may not be well-formed natural language sentences), whereas for sentence similarity tasks, the inputs are usually of comparable length and represent well-formed natural language.

I argue that these differences are primarily features of the annotation process for the training data and are captured in parametric variations of the logical scoring model defined in Eq.~(\ref{eq:logical}).
In practical terms, these task distinctions affect implementation design choices.
Is the relation we're trying to model symmetric?
In that case, let's just use the same encoder for both inputs.
Otherwise, having separate encoders makes more sense.
Interestingly, results from the dense retrieval model ANCE~\citep{Xiong_etal_ICLR2021}, which uses the same encoder for both queries and documents (despite obvious differences between the inputs), has been shown to work well empirically.
Maybe these design choices aren't so important anyway?

The goal of this discussion is to illustrate that the conceptual framework proposed in this paper establishes connections between information retrieval and natural language processing, with the hope that these connections can lead to further synergies in the future.
\citet{Lin_etal_2021_ptr4tr} (Chapter 5) argued that until relatively recently, solutions to the text retrieval problem and sentence similarity tasks have developed in relative isolation in the IR and NLP communities, respectively, despite the wealth of connections.
In fact, both communities have converged on similar solutions in terms of neural architectures (in the pre-BERT days).
The proposed conceptual framework here makes these connections explicit, hopefully facilitating a two-way dialogue between the communities that will benefit both.

\subsection{Historical Connections}

Civilizations have grappled with the challenges of accessing stored information shortly after the invention of writing, when humankind's collective knowledge outgrew the memory of its elders.
We can imagine some ancient scribe, perhaps somewhere in Mesopotamia, scrawling on clay tablets, wondering where he\footnote{As yes, very likely a male.} put those records from last month.
Libraries and archives, of course, have existed for millennia, created precisely to tackle this challenge.
In contrast, our conceptualization of information retrieval using computers is less than a century old.
Although the technologies have evolved over millennia, from clay tablets to scrolls to books, and now digital information, the underlying goals have changed little.

Interestingly, it is possible to apply the conceptual framework proposed in this paper to describe information retrieval in the eras that pre-dated computers.
For centuries, human librarians have been assigning content descriptors to information objects (books, scientific articles, etc.).
These descriptors (also known as ``index terms'') were usually selected by human subject matter experts and drawn from thesauri, ``subject headings'', or ``controlled vocabularies''---that is, a predefined vocabulary.
This process was known as ``indexing'' or ``abstracting'';
the original sense of the activity involved humans, and thus, an indexer was a human who performed indexing, not unlike the earliest uses of computers to refer to humans who performed computations by hand!
In other words, a human indexer served the role of the document encoder $\eta_d$, and the output can be viewed as a multi-hot vector where each of the dimensions represents a content descriptor.

Searching required the assistance of librarians who ``interviewed'' the information seeker to understand the parameters of the request, to translate the information need into the same representational space of these content descriptors.
Thus, librarians served the role of the query encoder $\eta_q$.
What about $\phi$?
Since the query and document representations are best characterized as multi-hot vectors, representation matching occurs in a boolean fashion.

In fact, the logical/physical separation applies to this human-mediated approach as well!
To ``execute'' retrieval in the simplest case of one-hot representations of content descriptors, the librarian consults a guide that maps these content descriptors into physical shelf locations, and then walks with the information seeker directly over to that location.
More sophisticated physical retrieval models include the use of card catalogues.\footnote{Millennials and even younger readers ask, ``What are those?''}
In the early days of computing, $\phi$ was implemented via the processing of punch cards,\footnote{Anyone other than boomers asks, ``What are those?''} each of which encoded the representation of an information object (i.e., the output of the document encoder $\eta_d$).
Thus, as a bonus, the conceptual framework proposed in this paper can help us understand information retrieval through the ages, even prior to the advent of computing.

\section{Experimental Results}
\label{section:exp}

\begin{table*}
\centering
\begin{small}
\begin{tabular}{lllrl}
\toprule
\multicolumn{3}{l}{\bf Unsupervised Sparse Representations} & {\bf MRR@10} & {\bf Source} \\
\toprule
(1) & \multicolumn{2}{l}{BM25} & 0.184 & \cite{Nogueira_Lin_docTTTTTquery} \\[1ex]
\toprule
\multicolumn{3}{l}{\bf Learned Sparse Representations} & {\bf MRR@10} & {\bf Source} \\
\toprule
& Term Weighting & Expansion \\
\midrule
(2) & BM25 & doc2query--T5 & 0.277 & \cite{Nogueira_Lin_docTTTTTquery} \\
(3) & DeepCT  & None &  0.243 & \cite{Dai:1910.10687:2019}\\
(4) & SparTerm & MLM-based & 0.279 & \cite{Bai:2010.00768:2020} \\
(5) & DeepImpact  & doc2query--T5 & 0.326 & \cite{Mallia_etal_SIGIR2021} \\[0.3ex]
\cdashline{1-5}\noalign{\vskip 0.5ex}
(6a) & COIL-tok ($d=32$)  & None & 0.341 & \cite{gao-etal-2021-coil} \\
(6b) & COIL-tok ($d=32$) & doc2query--T5 & 0.361 & \cite{Lin_Ma_arXiv2021} \\[0.3ex]
\cdashline{1-5}\noalign{\vskip 0.5ex}
(7a) & uniCOIL  & None & 0.315 & \cite{Lin_Ma_arXiv2021} \\
(7b) & uniCOIL  & doc2query--T5 & 0.352 & \cite{Lin_Ma_arXiv2021} \\
(7c) & uniCOIL  & TILDE & 0.349 & \cite{Zhuang_Zuccon_SIGIR2021} \\[0.3ex]
\cdashline{1-5}\noalign{\vskip 0.5ex}
(8a) & SparTerm/SPLADE & none & 0.290 & \cite{Formal_etal_SIGIR2021} \\
(8b) & SPLADE & MLM-based & 0.322 & \cite{Formal_etal_SIGIR2021} \\
(8c) & DistilSPLADE-max & MLM-based & 0.368 &\cite{Formal:2109.10086:2021} \\[1ex]
\toprule
\multicolumn{3}{l}{\bf Learned Dense Representations} & {\bf MRR@10} & {\bf Source} \\
\toprule
(9) & ColBERT & & 0.360 & \cite{Khattab_Zaharia_SIGIR2020} \\
(10) & ANCE & & 0.330 & \cite{Xiong_etal_ICLR2021} \\
(11) & DistillBERT &  & 0.323 & \cite{Hofstatter:2010.02666:2020} \\
(12) & RocketQA & & 0.370 & \cite{qu-etal-2021-rocketqa} \\
(13) & TAS-B & & 0.347 & \cite{Hofstatter_etal_SIGIR2021} \\
(14) & ADORE + STAR & & 0.347 & \cite{Zhan_etal_SIGIR2021} \\
(15) & TCT-ColBERTv2 & & 0.359 & \cite{Lin_etal_2021_RepL4NLP} \\[1ex]
\toprule
\multicolumn{3}{l}{\bf Dense--Sparse Hybrids} & {\bf MRR@10} & {\bf Source} \\
\toprule
(16) & CLEAR & & 0.338 & \cite{Gao_etal_ECIR2021_CLEAR} \\
(17) & COIL-full  & & 0.355 & \cite{gao-etal-2021-coil} \\
(18a) & \multicolumn{2}{l}{TCT-ColBERTv2 (15) + BM25 (1)} & 0.369 & \cite{Lin_etal_2021_RepL4NLP} \\
(18b) & \multicolumn{2}{l}{TCT-ColBERTv2 (15) + doc2query--T5 (2)} & 0.375 & \cite{Lin_etal_2021_RepL4NLP}  \\
(18c) & \multicolumn{2}{l}{TCT-ColBERTv2 (15) + DeepImpact (5)} & 0.378 & \cite{Lin_Ma_arXiv2021} \\
(18d) & \multicolumn{2}{l}{TCT-ColBERTv2 (15) + uniCOIL (7b)} & 0.378 & \cite{Lin_Ma_arXiv2021} \\
\bottomrule
\end{tabular}
\end{small}

\caption{Results on the development queries of the MS MARCO passage ranking task.}
\label{table:results}
\end{table*}

We can apply the conceptual framework proposed in this paper to organize various dense and sparse retrieval methods that have been proposed in the literature.
This structure can facilitate comparisons across different classes of methods, and analyzing models in a common framework can perhaps help us better draw generalizations.
Table~\ref{table:results} shows the effectiveness of various models on the development queries of the MS MARCO passage ranking test collection~\citep{MS_MARCO_v3}, which has emerged in recent years as the most prominent dataset for training and benchmarking retrieval models.

As a baseline, row (1) shows the effectiveness of BM25, which can be characterized as an unsupervised sparse retrieval method.
Learned sparse retrieval methods are shown in the second main block of Table~\ref{table:results}, from row (2) to row (8c):\
per the discussion in Section~\ref{section:logical-physical}, I break out term weighting and document expansion components.
BM25 with doc2query--T5 document expansions~\citep{Nogueira_Lin_docTTTTTquery}, row (2), can be understood as using a neural sequence-to-sequence model for expansion, but retaining the BM25 weighting scheme; thus, learning is only applied in the expansion component.
DeepCT~\citep{Dai:1910.10687:2019}, row (3), uses a regression-based term weighting model without any expansion.
SparTerm~\citep{Bai:2010.00768:2020}, row (4), uses the masked language model (MLM) layer of BERT to generate expansion terms on which term weights are learned.
DeepImpact~\citep{Mallia_etal_SIGIR2021}, row (5), combines the use of doc2query--T5 for expansion with a term weighting model trained using pairwise loss.

Rows (6a) and (6b) present a contrastive condition comparing the same term weighting model---COIL \citep{gao-etal-2021-coil}---with and without an expansion model; adding document expansion yields a two-point gain in effectiveness.
With uniCOIL~\citep{Lin_Ma_arXiv2021}, which builds on COIL, the literature reports three contrastive conditions:\ without expansion, row (7a), and with two different expansion methods, doc2query--T5 in row (7b) and TILDE~\citep{Zhuang_Zuccon_SIGIR2021} in row (7c).
These results affirm the importance of document expansion, but suggest that the exact choice of the model might not matter so much, at least in the uniCOIL design, since the expansion model simply provides a candidate list of terms for the term weighting model to consider during training.
Finally, row group (8) reports the effectiveness of a family of models called SPLADE, v1~\citep{Formal_etal_SIGIR2021} and v2~\citep{Formal:2109.10086:2021}, both of which build on SparTerm~\citep{Bai:2010.00768:2020}.
These results corroborate the importance of term expansions in learned sparse representations.

In the third main block of Table~\ref{table:results}, I summarize the effectiveness of a number of learned dense retrieval models on the development queries of the MS MARCO passage ranking test collection.
Note that ColBERT~\citep{Khattab_Zaharia_SIGIR2020} uses the more expressive MaxSim operator to compare query and document representations (more discussion in Section~\ref{section:discussion}); all other models use inner products.
Comparing dense vs.\ sparse learned representations, there does not appear to be any discernible pattern that can be identified.
While earlier proposals for learned sparse models under-perform learned dense models, it is likely because researchers have been investigating learned dense representations for a longer period of time.
From the perspective of effectiveness, the latest dense and sparse methods appear to be on par with each other.

The final block of Table~\ref{table:results} shows the results of dense--sparse hybrids.
In particular, rows \mbox{(18a--d)} present results of the TCT-ColBERTv2 dense retrieval model~\citep{Lin_etal_2021_RepL4NLP} with different learned sparse retrieval models using a simple linear combination of scores.
The only point I wish to make here is that dense and sparse representations appear to offer complementary relevance signals, such that combining evidence from both sources yields further increases in effectiveness compared to ranking with each individually.
However, it appears that hybrid fusion is less sensitive to the effectiveness of the individual models---for example, DeepImpact is less effective than uniCOIL, but both achieve the same effectiveness in a fusion context, as shown in row (18c) vs.\ row (18d).
Furthermore, fusion with doc2query--T5 achieves nearly the same level of effectiveness, shown in row (18b), even though the method alone is far less effective.
Overall, I believe that dense--sparse hybrids represent the state of the art in single-stage retrieval models today (i.e., what can be achieved without reranking).

\section{Discussion}
\label{section:discussion}

The conceptual framework described in this paper clarifies the relationship between recently proposed dense and sparse retrieval methods, and experimental results presented in the previous section begin to help us understand the impact of different design choices.
Furthermore, this proposed framework suggests a number of open research questions, which provide a roadmap for future work.
I discuss these below:

\paragraph{Out-of-distribution inference}
In the logical scoring model, explicitly establishing a contrast between supervised (learned) vs.\ unsupervised representations makes it obvious why DPR is more effective than BM25.
However, in a supervised machine-learning paradigm, we are immediately led to the obvious follow-up question:\ What happens if the trained models are applied to out-of-distribution data?
Phrased differently, what is the effectiveness of learned representations in a zero-shot setting?
Cast into the same parlance for comparison purposes, BM25 is always applied in a ``zero-shot'' manner (although admittedly, such a statement sounds odd).

In the information retrieval context, since training data typically comprise (query, relevant document) pairs, out of distribution could mean a number of different things:\ (1) the document encoder is fed text from a different domain, genre, register, etc.\ than the training documents, (2) the query encoder is fed queries that are different from the training queries, (3) the relationship between input query--document pairs at inference time differs from the relationship captured in the training data (e.g., task variations), or (4) a combination of all of the above.

In fact, we already know the answer, at least in part:\ learned representations often perform terribly in out-of-distribution settings when applied in a zero-shot manner.
Evidence comes from the BEIR benchmark~\citep{Thakur:2104.08663:2021}, which aims to evaluate the effectiveness of dense retrieval models across diverse domains.
Results show that, in many cases, directly applying a dense retrieval model trained on one dataset to another dataset sometimes yields effectiveness that is worse than BM25.
Complementary evidence comes from~\citet{Li_etal_FindingsEMNLP2021}, who found that for passage retrieval in question answering, training DPR on one dataset and testing on another can lead to poor results.
In their experiments, the corpus was fixed (Wikipedia articles), but the questions are generated in different ways; the end result is that the trained encoders often generalize poorly across datasets.

In contrast to BM25, which ``just works'' regardless of the corpus and queries in a ``zero-shot'' manner, learned representations may perform poorly in out-of-distribution settings.
This immediately suggests one important research direction, to better cope with these issues.
For example, \citet{Li_etal_FindingsEMNLP2021} proposed model uncertainty fusion as a solution.
The BEIR benchmark~\citep{Thakur:2104.08663:2021} provides a resource to evaluate progress, and the latest results show that learned sparse representations are able to outperform BM25~\citep{Formal:2109.10086:2021}.
At a high level, there are at least three intertwined research questions:

\begin{enumerate}[leftmargin=0.75cm]

\item What are the different ways in which models can be applied in an out-of-distribution manner and what is the impact of each?
The four ways I've sketched above provide a starting point, but could be further refined with experimental support.
For example, is effectiveness degradation more severe with out-of-distribution documents or queries?
Can we more formally characterize ``out-of-distribution''--ness?

\item Given the answers to the above questions, how do we then detect when an input instance is out of distribution?

\item And once we identify a potentially ``problematic'' instance, what mitigation techniques can we bring to bear?

\end{enumerate}

\noindent In other words, we must understand the scope of the problem, identify when the problem occurs, and then finally mitigate the problem.
Without addressing these challenges, the real-world deployment of learned representations will be hampered by their inability to generalize to arbitrary information retrieval scenarios, in the way that BM25 isn't.\footnote{Another way to say this:\ Suppose we're faced with a completely new retrieval task in a highly specialized and obscure domain. I think most researchers and practitioners would unequivocally suggest using BM25 as the baseline, and would be confident of obtaining ``reasonable'' results. I don't think we have that same confidence with any learned representations at present.}
I am heartened to see that the community has already begun to explore these interesting and important research questions, but there remains much more work to be done.

\paragraph{Quality--Space--Time--Cost tradeoffs}
By situating dense and sparse retrieval models in a unified conceptual framework, comparisons between different methods become more meaningful.
There are four dimensions along which different retrieval models should be compared:\ quality (e.g., retrieval effectiveness), space (e.g., index size), time (e.g., query latency), and cost (e.g., dollars per query).
Naturally, most papers today focus on output quality, but the space requirements of dense vector representations have drawn interest from researchers as well.

Retrieval models that depend on dense vector representations consume a large amount of space, which often translates into large memory requirements since many approximate nearest neighbor search libraries require memory-resident index structures for efficient querying.
For example, a minimal Lucene index in Anserini~\citep{Yang_etal_JDIQ2018}, sufficient to support bag-of-words querying on the MS MARCO passage corpus (8.8M passages), takes up only around 660 MB.
A comparable HNSW index with 768-dimensional vectors in Faiss occupies 42 GB (with typical parameter settings), which is many times larger.
As another example, \citet{Ma_etal_EMNLP2021} reported that the size of the original DPR (flat) vector index on the Wikipedia corpus is about 61 GB,\footnote{An HNSW index suitable for low-latency querying would be even larger.} compared to 2.4 GB for a comparable Lucene inverted index.
This $25\times$ increase in space only yields an average gain of $2.5\%$ in top-$100$ accuracy across five datasets~\citep{Ma2021ARS}.

While researchers have begun to explore different techniques for reducing the space requirements for dense representations, for example, via dimensionality reduction or quantization~\citep{Izacard:2012.15156:2020,yamada-etal-2021-efficient,Ma_etal_EMNLP2021}, there is much more work to be done.
I am optimistic that the community will make headway here because, as already mentioned above, the comparisons to sparse representations are ``not fair'' because inverted indexes have benefited from many decades of optimizations, particularly in the coding of sparse integer sequences, whereas researchers have only begun to tackle the impractically large space requirements associated with dense retrieval models.

Finally, speed (more generally, performance characterized in terms of query latency, throughput, etc.)~and cost (of hardware, power consumption, amount of CO$_2$ generated, etc.)~are issues that have received comparatively little attention, but are obviously important in real-world applications.
I mention these considerations in tandem because there are many examples where, holding everything else fixed, speed and cost can be traded off for each other.
A simple example is GPU vs.~CPU inference for retrieval models that require neural inference on queries, which must be performed at search time.
Since queries are usually short, CPU inference, even with transformer models, can be tolerable, but obviously, GPU inference can reduce query latency but incur additional hardware costs.
As another example, in many real-world search applications, query latency can be controlled in partitioned architectures by adjusting the size of each partition (also called a shard):\ the smaller each partition, the lower the query latency, but at the cost of needing more hardware (and hence cost) for a given corpus size.
While there have been some discussions of these issues in blog posts\footnote{For example, the ``Pretrained Transformer Language Models for Search'' series at {\small \url{https://blog.vespa.ai/}}.} and on social media, these considerations have not attracted much attention from researchers.

Moving forward, I believe that an accurate characterization of dense and sparse retrieval methods requires clearly evaluating quality--space--time--cost tradeoffs.
This to me is exciting because it provides an opportunity for collaborations between ``modeling--minded'', ``algorithm--minded'', and ``efficiency--minded'' researchers.\footnote{More colloquially, our colleagues who get their kicks reducing L1 cache misses and bits per posting can now get in on the neural action.}

\paragraph{``Mixing and matching'' logical scoring models and physical retrieval models}
Dense and sparse representations are not discrete categories, but rather lie on a continuum with many variations.
Currently, the size (in terms of the number of dimension) of (most) sparse representations equals the vocabulary size of the corpus, and dense representations typically have hundreds of dimensions (768 being a common setting).
What if we ``densify'' sparse representations and ``sparsify'' dense representations---to yield, say, vectors that are on the order of a few thousand dimensions?
We might characterize these vectors as ``not really dense, but not sparse either''.
For such a logical scoring model, what physical retrieval model makes the most sense in terms of different tradeoffs?

In Section~\ref{section:logical-physical}, I advocated for the separation of the logical scoring model from the physical retrieval model.
A loosely coupled approach provides flexibility and the ability to make progress independently on different aspects of the overall problem.
Currently, there is an affinity between sparse representations and query evaluation using inverted indexes on the one hand, and dense representations and HNSW on the other.
But what happens when the representations move out of their respective ``sweet spots''?
As we ``densify'' sparse representations, the performance of inverted indexes is expected to degrade.
As we ``sparsify'' dense representations, the performance of HNSW is expected to degrade.
Thus, we expect some crossover point in the middle?
Perhaps for vectors that are ``not really dense, but not sparse either'', neither approach will work well.
This suggests a need to build index structures coupled with algorithmic innovations for top-$k$ retrieval on such vector representations.

I believe that this is where a clean abstraction and the ability to ``mix and match'' different logical scoring models with physical retrieval models will really become beneficial.
We can imagine the development of different data structures and algorithms targeted to different types of representations---beyond the (basically, two) limited options we have today.
Depending on the characteristics of the vector representations, for example, the number of dimensions, the entropy of the values, the degree of isotropy, etc., different physical retrieval models might be appropriate.
This is taking a page out of the playbook of database researchers---for example, it is precisely the logical/physical abstraction that has enabled the development of very different types of database engines such as row stores and column stores for different application scenarios~\citep{Stonebraker_etal_VLDB2005}.
And who knows, maybe we can even {\it learn} physical retrieval models~\citep{Idreos_etal_2019}!

\paragraph{Alternative comparison functions}
For both sparse and dense representations, the inner product holds a privileged position as the comparison function $\phi$ because efficient solutions already exist for the top-$k$ retrieval problem.
As I already explained in Section~\ref{section:logical-physical}, fixing $\phi$ to be the inner product allows a researcher to focus on the logical scoring model in isolation (notwithstanding the issues discussed above).
This is a good compromise because limiting $\phi$ to be the inner product still leaves open the entire space of neural architectures for designing the encoders---and indeed, most dense retrieval research operates under this constraint.

The framework does not, however, preclude alternative definitions of $\phi$---rather, it just means that a ``custom'' comparison function may need its own dedicated physical retrieval model (unless, that is, we solve the challenges discussed above).
A good example is ColBERT~\citep{Khattab_Zaharia_SIGIR2020}, which introduced a comparison function called ``MaxSim'' that computes query--document similarity as the sum of the maximum cosine similarities between each query term and the ``best'' matching document term; cf.~\citet{Kusner_etal_ICML2015}.
To efficiently compute top-$k$ rankings in terms of MaxSim, the authors first built an index for approximate nearest neighbor search over {\it all tokens} in the document collection, where each token retains a pointer back to its source document.
Retrieval is performed by first fetching candidate documents using this index (by following the pointers) and then computing MaxSim for all query--document candidates.
In other words, the authors presented a two-stage physical retrieval model specifically for their novel comparison function.

In fact, ColBERT offers a good example where many of the discussion threads above come together.
Khattab and Zaharia described a design where the logical scoring model and the physical retrieval model are tightly coupled.
Separating the two might accelerate future advances by enabling independent progress.
On the one hand, researchers could rely on MaxSim as $\phi$ and explore different query or document encoders without worrying about retrieval efficiency.
On the other hand, another group of researchers could focus on optimizing MaxSim calculations over large document collections without worrying about whether such optimizations would be useful.
In this way, MaxSim might gain a ``privileged'' status, alongside the inner product, in the selection of the comparison function $\phi$ for retrieval model design.

In addition, ColBERT provides an illustrative case study for the need to characterize quality--space--time--cost tradeoffs in order to compare retrieval models in a ``fair'' manner.
Khattab and Zaharia presented their innovation as a model that is just as effective as a retrieve-then-rerank approach using BERT-based cross-encoders~\citep{nogueira2019passage}, but is substantially faster.
This, however, comes at the cost of huge index sizes---154 GB for the MS MARCO passage corpus (compared to 660 MB for an inverted index).
While the authors did discuss this limitation, when all four dimensions of evaluation are considered (quality, space, time, and cost), it is difficult to see ColBERT as a practical solution for real-world problems.

\paragraph{Multi-stage ranking as physical optimizations}
In Section~\ref{section:logical-physical}, I argued that multi-stage ranking architectures are simply {\it practical} implementations of expensive logical scoring models (based on brute-force scans).
Here, I elaborate on this observation, which also bolsters the case for logical/physical separation.

Any multi-stage ranking pipeline where the scores from each stage are additive can be converted into the functional form of Eq.~(\ref{eq:logical}) by ``composing'' the models at each stage (including first-stage retrieval).
In a ranking pipeline where the later stages do not incorporate evidence from the earlier stages (that is, stages are used only to reduce the candidates under consideration), such as BM25 + monoBERT~\citep{nogueira2019passage}, the score of the final reranking stage {\it is} the logical scoring model.
In either case, top-$k$ retrieval can be performed using a brute-force scan through the entire document collection based on the logical scoring model.
Thus, multi-stage pipelines can be viewed as {\it hand-crafted} optimizations in the physical retrieval model.

In other words, with a clean logical/physical separation, researchers and practitioners can focus on developing the logical scoring model, leaving the realization of the physical retrieval model as a separate exercise.
In the tightly coupled architectures of today, the logical scoring model and the physical retrieval model must be co-designed to produce the ``right'' multi-stage pipeline.
This is inelegant, as designers are mixing elements from different levels of abstraction:\ {\it what} to compute with {\it how} to compute.
However, this conceptual tangle need not be the only approach.
For example, we might build automated processes that ``compile'' the specification of the logical scoring model into a physical realization, subjected to declaratively specified constraints.
These hypothetical logical-to-physical compilers can even be machine learned!
The work of~\cite{Wang_etal_SIGIR2011} provides an example of how this could be accomplished in the context of feature-based learning to rank; perhaps these ideas from a decade ago could be dusted off for a fresh take?

\paragraph{Unsupervised dense representations}
The conceptual framework proposed in this paper characterizes logical scoring models along two dimensions.
The four-quadrant taxonomy illustrated in Table~\ref{table:framework} highlights a space that has not received much attention of late.
I don't have much to say here, except that perhaps this gap might suggest a research direction worth renewed investigation.

\paragraph{Other odds and ends}
If the logical scoring model and the physical retrieval model represent abstractions that are helpful in advancing IR research, what other such abstractions might exist?
And a related question:\ So far, the conceptual framework proposed here has been applied primarily to deepen our understanding of {\it ad hoc} retrieval.
What, if any, implications does this framework hold for other areas of information seeking beyond the design of retrieval models?

Addressing the first question: An important abstraction that immediately comes to mind, although hardly novel, is that of a token stream as the input to an inverted indexer (and correspondingly, to a query processor prior to retrieval).
That is, an inverted indexer merely requires a stream of discrete tokens on which to operate, and is agnostic with respect to how the tokens are generated from arbitrary natural language text.
In the canonical case, these tokens correspond to ``words'' in the language (however defined) after some amount of analysis (e.g., stemming), but researchers have discovered, that at least for some languages, character $n$-grams (which have no basis in linguistic reality) work well~\citep{Foo04,McNamee_Mayfield_IRJ2004}.
Much along the same lines, \cite{Xue:2105.13626:2021} recently explored pretrained neural sequence-to-sequence models based on byte sequences and showed that such models are competitive to token-based models, but more robust to noisy inputs.
Perhaps it is worth reconsidering the information retrieval tokenization pipeline in light of these latest results?

Addressing the second question on whether the conceptual framework presented in this paper has anything meaningful to say about other areas of information retrieval and information seeking more broadly:
I can think of two answers.

First, it has long been observed that information filtering and {\it ad hoc} retrieval are intimately related, what~\citet{Belkin_Croft_CACM1992} have called ``two sides of the same coin''.
At a high level, {\it ad hoc} retrieval is concerned with a stream of queries posed against a (relatively) static collection of documents, whereas information filtering is concerned with a stream of documents posed against a (relatively) static collection of queries.
Filtering has a long history that dates back to the 1960s~\citep{Housman_Kaskela_1970}, which evolved into the TREC Filtering Tracks~\citep{Lewis_TREC1995} in the late 1990s and the general research program
known as Topic Detection and Tracking (TDT)~\citep{Allan_2002} in the early 2000s.
The most recent incarnations of filtering include the TREC Incident Streams Tracks~\citep{TRECIS-2020}, which aim to automatically process social media streams during emergency situations to triage information and aid requests for emergency service operators.
This evaluation series has its roots in the TREC Real-Time Summarization Tracks~\citep{Lin_etal_TREC2016}, where systems automatically monitor streams of social media posts to keep users up to date on topics of interest.

I believe that a more succinct way to convey the connections between filtering and {\it ad hoc} retrieval (cf.~\citet{Belkin_Croft_CACM1992}) is to say that they share logical scoring models---at least in terms of Eq.~(\ref{eq:logical}), although the relevance criteria are often different---but may require different physical retrieval models.
Although information filtering, in fact, can be physically implemented via inverted indexes, such a realization can be somewhat awkward (a side effect of the ``tight coupling'' approach).
A clean separation between the logical and physical can help researchers focus on representations and scoring models without artificial constraints on execution.
More clearly-defined sub-problems, I believe, will lead to accelerated progress in the field, with all the advantages I've already discussed above.

Second, I believe that the conceptual framework proposed here can capture relevance feedback (pseudo- or based on human judgments), and more generally, interactive retrieval.
The logical scoring model as currently defined computes the query representation from the query itself, i.e., $\eta_q(q)$.
However, this formalism can be extended to take into account previous queries in a session, e.g., $\eta_q(q_i; q_{<i})$, where $q_i$ denotes the query at the $i$-th turn, and $q_{<i}$ denotes all queries that came before.
This can further be extended to include the results of those previous queries, along with human input, e.g., relevance judgments.
In fact, most participants in the TREC Conversational Assistance Tracks~\citep{Dalton_etal_TREC2019_CaST}, as well as the design of conversational question answering~\citep{choi-etal-2018-quac,elgohary-etal-2019-unpack}, adopt this formulation, either implicitly or explicitly.
This suggests that interactive retrieval can be incorporated into the conceptual framework proposed here with appropriate extensions.
No doubt there is much more work to be done and there may be other aspects of information seeking that do not neatly fit into the existing formalisms, but I believe that this exposition provides a helpful start.

\section{Conclusions}

The conceptual framework for text retrieval presented in this paper allows us to ``tie together'' recent work in dense and sparse retrieval, along with multi-stage ranking architectures, from a representational perspective that identifies the logical scoring model and the physical retrieval model as the ``right'' abstractions.
We can understand recent developments not as disjoint innovations, but different aspects of the same underlying research program that has remained obscured until now.
This understanding lets us look forward, by suggesting a number of open research questions that provide a roadmap for future advances, look backward, by establishing historic connections to information access ``technologies'' dating back millennia, and look around, by enhancing linkages to natural language processing and other areas of information retrieval.
With so much ``going on'' and rapid advances at every turn, it is an exciting time for information retrieval research!

\section*{Acknowledgements}

This research was supported in part by the Natural Sciences and Engineering Research Council (NSERC) of Canada.
The ideas presented here have benefited from valuable discussions with Xueguang Ma, Joel Mackenzie, Antonio Mallia, Rodrigo Nogueira, Andrew Yates, and Shengyao Zhuang.
Thanks to Jo Kristian Bergum for pointing out that speed should be considered in the context of cost, as the two can be traded off.
Additional thanks to the audience at DESIRES 2021, where the presentation and ensuing discussion of an earlier rendition of these ideas~\citep{Lin_etal_DESIRES2021} further refined my thinking, as captured in ``other odds and ends'' in Section~\ref{section:discussion}.

\bibliographystyle{abbrvnat}
\bibliography{rep_ir}

\end{document}